%%%%%%%%%%%%%%%%%%%%%%%%%%%%%%%%%%%%%  FONTI

\font\elvrm=cmr10  scaled\magstep1
\font\elvmi=cmmi10  scaled\magstep1
\font\elvsy=cmsy10  scaled\magstep1
\font\elvex=cmex10  scaled\magstep1
\font\elvit=cmti10  scaled\magstep1
\font\elvsl=cmsl10  scaled\magstep1
\font\elvbf=cmbx10  scaled\magstep1

\font\ninerm=cmr9         \font\eightrm=cmr8          \font\sixrm=cmr6
         \font\eighti=cmmi8          \font\sixi=cmmi6
        \font\eightsy=cmsy8         \font\sixsy=cmsy6

\newcount\textno
\newcount\scriptno
\global\textno=10
\global\scriptno=7

%%%%% SCELTA FONTI - CANCELLARE I BLOCCHO CHE NON INTERESSANO

%%%%%%%%%%%%%%%%%%%%%%%%%%%%%%%%%%%%%%FONTI A 12 PT  ELVPOINT
\textno=12\scriptno=7
\normalbaselineskip=13dd  % \normallineskip=4pt \normallineskiplimit=4pt
\textfont0=\elvrm  \scriptfont0=\eightrm  \scriptscriptfont0=\sixrm
\textfont1=\elvmi  \scriptfont1=\eighti  \scriptscriptfont1=\sixi
\textfont2=\elvsy  \scriptfont2=\eightsy  \scriptscriptfont2=\sixsy
\textfont3=\elvex  \scriptfont3=\elvex    \scriptscriptfont3=\elvex
\textfont4=\elvit  \textfont5=\elvsl      \textfont6=\elvbf
\def\rm{\fam0\elvrm}                      \def\it{\fam4\elvit}
\def\sl{\fam5\elvsl}                      \def\bf{\fam6\elvbf}
\setbox\strutbox=\hbox{\vrule height8.5dd depth4dd width0dd}
\normalbaselines\rm
%LOCALIZZAZIONE FILE
{\vskip-2cm
\headline={\tt %%NUMERO DISK
$\backslash$ %%%NOME DIR
$\backslash$ %%%NOME FILE
}\vskip2cm}

%%%%%%%%%%%%%%%%%%%  INTESTAZIONE
\nopagenumbers
\def\rightheadline{{\eightrm\hfil
%% SCRIVERE QUI L'INTESTAZIONE DELLA PAGINA DISPARI E CANCELLARE I %%
\hfil}\folio}
\def\leftheadline{\folio\eightrm\hfil
%% SCRIVERE QUI L'INTESTAZIONE DELLA PAGINA PARI E CANCELLARE I %%
\hfil}
\headline={\vbox to 0pt{\vskip-1.82pc\line{\vbox to 8.5pt{}
\ifodd\pageno\rightheadline \else\leftheadline\fi}\vss}}
%%%%%%%%%%%%%%%%%%%%%%%%%%%%%%%%%%%%%%%%%%%%%

%%%%%%%%%%%%%%%%%%%%%%%%%%%%FORMATO
\vsize19.2cm \hsize12.8cm \baselineskip=6.4mm
\voffset2.5pc\hoffset2.5pc

%%%%%%%%%%%%%%%%%%%  INTESTAZIONE
\nopagenumbers
\def\rightheadline{{\eightrm\hfil
%% SCRIVERE QUI L'INTESTAZIONE DELLA PAGINA DISPARI E CANCELLARE I %%
\hfil}\folio}
\def\leftheadline{\folio\eightrm\hfil
%% SCRIVERE QUI L'INTESTAZIONE DELLA PAGINA PARI E CANCELLARE I %%
\hfil}
\headline={\vbox to 0pt{\vskip-1.82pc\line{\vbox to 8.5pt{}
\ifodd\pageno\rightheadline \else\leftheadline\fi}\vss}}
%%%%%%%% PARTICULAR MACROS
%%%%%%%%%%%%%%%%%%%%%%%%%%
\centerline{\bf Individual consistency}
\par
\centerline{\bf of 2-events quantum histories}
\vskip1.2cm
\centerline{Giuseppe Nistic\`o}
\par
\centerline{\it Dipartimento di Matematica -
Universit\`a della Calabria}
\par
\centerline{87036 Arcavacata, Rende (CS), Italy}
\par
\centerline{gnistico@unical.it}
\vskip6mm
\centerline{Roberto Beneduci}
\par
\centerline{\it Dipartimento di Fisica -
Universit\`a della Calabria}
\par
\centerline{87036 Arcavacata, Rende (CS), Italy}
\par
\centerline{cabeneduci@libero.it}
\vskip4.6mm\noindent
{\ninerm 
It is argued that the property of consistency of
consistent history approach to quantum physics
is not an individual property, in the sense that when such consistency
holds, it cannot be attributed to each single sample of
the physical system. This fact is not a logical inconsistency
but it is in striking contrast with the physical idea of consistency.
In this letter we introduce a meaningful notion of consistency, named
self-decoherence, based on the concept of mirror projection.
We prove that self-decoherence satisfies our tentative
criterion of individuality.
Furthermore, it is proved that self-decoherence forbids contrary inferences.}
\vskip1.2cm\noindent
In 1984 R.B. Griffiths [1] proposed a reinterpretation of quantum formalism
with the aim of giving a solution to the ``well-known conceptual
difficulties which arise in various interpretations of
quantum mechanics''.
While standard quantum theory is based on the concept of {\sl event},
represented by a projection operator $E$ of the Hilbert space $\cal H$
describing the system, the {\sl consistent history approach} (CHA) is based
on the concept of {\sl history}, which is any finite ordered sequence
$h=(E_1,E_2,...,E_n)$ of events. The CHA provides the framework
in which it is possible to establish whether histories have physical
meaning [2]. Such framework is made up of suitable {\sl families}
of histories. Let
${\bf E}_1,{\bf E}_2, ...,{\bf E}_n$
be finite resolutions of the
identity, i.e ${\bf E}_k= \{E_k^{(1)},E_k^{(2)}, ... , E_k^{(i_k)} \}$,
where $E_k^{(i)}\perp E_k^{(j)}$ if $i\neq j$ and
$\sum_{i=1}^{i_k}E_k^{(i)}={\bf 1}$.
%(they could be considered as the spectral resolutions of $n$
%quantum observables, not necessarily commuting).
A {\sl family} $\cal C$ of histories is the set of all histories
$h=(E_1,E_2,...,E_n)$ such that $E_k=\sum_{\hbox{\sevenrm some
}i}E_k^{(i)}$ for a fixed $n$-uple ${\bf E}_1,{\bf E}_2, ...,{\bf E}_n$
of resolutions of the identity. When every event $E_k$ constituting
a history $h$ is just an event of ${\bf E}_k$, i.e. if $E_k\in{\bf E}_k$
for all $k=1,2,...,n$, then $h$ is called {\sl elementary} history.
Hence the set $\cal E$ of all elementary histories of $\cal C$ is
the cartesian product ${\cal E}={\bf E}_1\times{\bf E}_2\times\cdots
\times{\bf E}_n$.
Two histories $h_1=(E_1,E_2,...,E_n),h_2=(F_1,F_2,...,F_n)\in\cal C$
are {\sl summable} if they
differ in only one place, say  $k$, hence $E_j=F_j$ for all $j\neq k$,
and $E_k\perp F_k$; in such a case their {\sl sum} is
$h_1+h_2=(E_1,E_2,...,E_k+F_k,...,E_n)\in\cal C$.
The histories $h_1$ and $h_2$ are said to be {\sl alternative}
if there is $k$ such that $E_k\perp F_k$.
\vskip4.6mm
Let $h=(E_1,E_2,...,E_n)$ be a {\sl commutative} history, i.e. all
$E_k$ commute with each other.
According to quantum theory, the statement ``$h$ {\sl occurs}'' means
that all events $E_1,E_2,...,E_n$ occur in the given order. Therefore,
$h$ is identified with the single event $E_1\cdot E_2\cdots E_n=
E_1\land E_2\land\cdots\land E_n$.
Though the mathematical notions of CHA are given within the standard
quantum theoretical
formalism, quantum theory is unable to consider and describe the
occurrence of a history when it is not commutative.
On the contrary, according to CHA, the histories of a family $\cal C$
have physical meaning whenever a condition of {\sl consistency} is
satisfied, which allows to assign a probability of occurrence $p(h)$
to every $h\in\cal C$. According to such idea of consistency, the
occurrence of an elementary history must imply the non-occurrence of
every other elementary history. Therefore, if there is a probability $p(h)$ of
occurrence of $h$, then it must satisfy
the {\sl sum rule}\par
$$
p\left(\sum_j h_j\right)=\sum_j p(h_j);\quad \sum_{h\in{\cal E}}p(h)=1.
\leqno(C.0)
$$
Moreover, the empirical validity of the theory requires that such
probability should be consistent with the probability assigned to
single events by quantum theory. Then, another condition for $p$
is
\item{(C.1)}
{\sl whenever $h=(E_1,E_2,...,E_n)$ and
$[E_j,E_k]={\bf 0}$ then
$$p(h)=Tr(E_nE_{n-1}\cdots E_1\rho),$$
where $\rho$ is the density operator such that $Tr(E\rho)$ is the
quantum probability of occurrence of the event $E$.}
\par\noindent
Condition (C.1) is satisfied if $p$ is
the functional $p:{\cal C}\to [0,1]$,
$p(h)=Tr(C_h\rho C_h^\ast)$, where
$C_h=E_nE_{n-1}\cdots E_1$. Such $p$ satisfies also (C.0)
if and only if [2]
$$
Re[Tr(C_{h_1}\rho C_{h_2}^\ast)]=0\quad\hbox{for all summable }
h_1,h_2\in{\cal E}.\eqno(1)
$$
When (1) holds, $\cal C$ is said to be {\sl weakly decohering}.
\par\noindent
D{\ninerm EFINITION} 1.
{\sl A family of histories $\cal C$ is said to be {\sl consistent} with
respect to $\rho$ if it is weakly decohering.}
\par
The following P1 and P2 are the basic principles of CHA.
\item{P1:}{\sl All predictions about the physical system are
those obtained by interpreting $p(h)=Tr(C_h\rho C_h^\ast)$
as probability of occurrence of $h$,
within a consistent family $\cal C$.}
\par\noindent
The notion of family of histories of CHA turns out to be
a generalization of the
notion of observable of standard quantum theory; this last can be
recovered within CHA by considering families of one-event histories
$h=(E)$,
i.e. generated by only one resolution of the identity.
As well as in standard quantum theory it is not possible to
non-contextually assign values to all observables [3], in CHA
it is not possible to assign the occurring histories in
{\sl all} consistent families together, without giving rise to {\sl contrary
inferences}, i.e. to contradictions of Kochen-Specker type [4].
This is the content of the {\sl single family}
rule:
\item{P2:}{\sl
the occurrence or the non-occurrence of a history $h$ can be considered
only within a single
consistent family $\cal C$, i.e. when $h\in\cal C$ and $\cal C$ is
weakly decohering.}
\par\noindent
The correct use of the basic principles of CHA makes it possible to recover
all results of standard quantum theory, avoiding important conceptual
difficulties [2].
\vskip1pc
The question we face in the present paper is whether the consistency
of a given family $\cal C$
is a property to be attributed to every single sample of the physical
system or not. From the point of view of our intuition, given
a consistent family $\cal C$ and a history $h\in\cal C$,
for each individual sample of the physical system there are two
mutually exclusive alternatives: either $h$ occurs or $h$ does not occur.
Therefore, the physical idea of consistency which is at the root of
CHA suggests that consistency should be an {\sl individual} property.
\par\noindent
In quantum theory and in CHA there are properties which are individual
and also properties which are not individual.
For instance, the property of having a given value $c$ of an
observable $C$ is individual. The following example shows that
the consistency of $\cal C$ in definition 1 is not an individual property.
\vskip4.6mm\noindent
E{\ninerm XAMPLE} 1. --
\noindent
Let us consider two density operators
$\rho_1=\vert\psi_1\rangle\langle\psi_1\vert$
and
$\rho_2=\vert\psi_2\rangle\langle\psi_2\vert$,
where $\psi_1$ and $\psi_2$ are two mutually orthonormal vectors of
$\cal H$.
Let $\varphi={1\over\sqrt 2}(\psi_1+\psi_2)$ be a third unit vector. If we put
$E_1=\vert\varphi\rangle\langle\varphi\vert$, then
$E_1\psi_1= {1\over 2}(\psi_1+\psi_2)=E_1\psi_2$
and $E_1'\psi_1= -E_1'\psi_2$, where $E_1'={\bf 1}-E_1$.
Therefore, taking $\rho={1\over 2}[\rho_1+\rho_2]$, we have
$$
Tr(E_2E_1\rho E_1^\prime E_2)
={1\over 2}[\langle E_1'\psi_1\mid E_2E_1\psi_1\rangle +
\langle E_1'\psi_2\mid E_2E_1\psi_2\rangle] =0
\eqno(2)
$$
for all projections $E_2$.
Since $Tr(E_2^\prime E_1\rho E_1 E_2)=0$ whatever $E_2$,
the family of histories $\cal C$ generated by the history
$(E_1,E_2)$ is consistent, whatever the projection operator $E_2$.
This $E_2$ can be chosen in such a way that
$\cal C$ turns out to be
consistent neither with respect to $\rho_1$,
nor with respect to $\rho_2$.
Indeed, by representing
vectors and operators of $\cal H$ with respect
to any fixed orthonormal basis $(u_n)_{n\in\bf N}$ so that
$u_1=\psi_1$ and $u_2=\psi_2$, we have
$\psi_1\equiv\left[\matrix{1\cr 0\cr {\bf 0}}\right]$,
$\psi_2\equiv\left[\matrix{0\cr 1\cr {\bf 0}}\right]$,
$E_1\equiv{1\over 2}\left[\matrix{1&1&{\bf 0}\cr 1&1&{\bf 0}\cr
{\bf 0}&{\bf 0}&{\bf 0}}\right]$.
Let us consider the histories
$h_1=(E_1,E_2)$, $h_2=({\bf 1}-E_1,E_2)$ and $h=h_1+h_2=({\bf 1},E_2)$,
where
$E_2\equiv\left[
\matrix{\cos^2{\theta\over 2}&-{i\over 2}\sin\theta&{\bf 0}\cr
{i\over 2}\sin\theta&\sin^2{\theta\over 2}&{\bf 0}\cr
{\bf 0}&{\bf 0}&{\bf 0}}\right], $
with $0<\theta<{\pi\over 2}$.
Then, $Tr(C_h\rho_1 C_h^\ast)=\cos^2{\theta\over 2}$, while
$Tr(C_{h_1}\rho_1 C_{h_1}^\ast)
=Tr(C_{h_2}\rho_1 C_{h_2}^\ast)={1\over 4}$, and this implies
$Tr(C_{h_1+h_2}\rho_1 C_{h_1+h_2}^\ast)\neq
Tr(C_{h_1}\rho_1 C_{h_1}^\ast)+
Tr(C_{h_2}\rho_1 C_{h_2}^\ast)$. The same argument applied to $\rho_2$
shows that
$Tr(C_{h_1+h_2}\rho_2 C_{h_1+h_2}^\ast)\neq
Tr(C_{h_1}\rho_2 C_{h_1}^\ast)+
Tr(C_{h_2}\rho_2 C_{h_2}^\ast)$. Therefore
the family $\cal C$ generated by $h_1$ and $h_2$ is {\sl not}
consistent
with respect to $\rho_1$ and $\rho_2$, but
it becomes consistent by mixing together the two statistical
ensembles represented by $\rho_1$ and $\rho_2$,
i.e. with respect to the mixture
$\rho={1\over 2}[\rho_1+\rho_2]$.
\vskip4.6mm
Example 1 suggests the following tentative definition of what is
an ``individual property'' of the physical system.
\vskip4.6mm\noindent
D{\ninerm EFINITION} 2 --
{\sl A property $\pi$ is individual for a quantum system
if the following statement holds.}
\item{\quad}
{\sl If $\pi$ does not hold when the system is described by $\rho_1$ or
$\rho_2$, then $\pi$ does not hold when the system is described by any
mixture $\rho=\lambda\rho_1+(1-\lambda)\rho_2$, with $0<\lambda<1$.}
\vskip4.6mm
It must be said that several notions of consistency other than
weak decoherence have been introduced in literature to achieve
a more strict adherence with the idea of consistency.
\par
M. Gell-Mann and J.B. Hartle [5] introduced
the stronger notion of {\sl medium decoherence}:
a family $\cal C$ has the property of medium decoherence if
$Tr(C_{h_1}\rho C_{h_2}^\ast)=0$ for all alternative $h_1,h_2\in\cal C$.
Now, from (3) it follows that the family $\cal C$ of example 1
has the property of medium decoherence with respct to $\rho$; but with
respect to $\rho_1$ and $\rho_2$ it is not weakly decohering
and therefore even medium decoherence does not hold.
Thus, medium decoherence is not an individual property.
\par
The {\sl linearly positive} decoherence proposed by S. Goldstein
and D.N. Page [6] consists in requiring that $Re[Tr(C_h\rho)]\geq 0$
for all $h\in\cal C$; it is weaker than weak decoherence. Therefore,
the family $\cal C$ of example 1 is also linearly positive with respect to
$\rho$, whatever $E_2$.
We can choose $E_2$ so that $\cal C$ is not linearly positive with
respect to $\rho_1$.
Let us consider the projection operator
$$
E_2=\left[\matrix{
\cos^2{\theta\over 2}&{1\over 2}e^{-i\alpha}\sin\theta&{\bf 0}\cr
{1\over 2}e^{i\alpha}\sin\theta&\sin^2{\theta\over 2}&{\bf 0}\cr
{\bf 0}&{\bf 0}&{\bf 0}\cr}\right],
$$
and the history $h_1=(E_1,E_2)$. We have
$$
Tr(C_{h_1}\rho_1)=\langle\psi_1\mid E_2E_1\psi_1\rangle=
{1\over 2}\left(\cos^2{\theta\over 2}+e^{-i\alpha}\sin{\theta\over 2}
\cos{\theta\over 2}\right).
$$
Therefore, for $0<\theta<{\pi\over 2}$ the condition
$Re(Tr[C_{h_1}\rho_1])\geq 0$ of linear positivity becomes
%Re(Tr[C_{h_1}\rho_1])={1\over 2}[\cos^2{\theta\over %2}+\cos\alpha\sin{\theta\over 2}
%\cos{\theta\over 2}].
%$
$\cos{\theta\over 2}+\cos\alpha\sin{\theta\over 2}\geq 0$ and it can be violated
by a suitable choice of $\theta$ and $\alpha$.
Thus, also linear positivity violates the individuality condition.
\par
Now we consider the {\sl ordered consistency} introduced by A. Kent
to avoid contrary inferences [4].
Following A. Kent we define the ordering
$
h_1\leq h_2$ iff
$E_k\leq F_k$ for all $k$, where $h_1=(E_1,E_2,...,E_k,...)$ and
$h_2=(F_1,F_2,...,F_k,...)$. A history $h_1$ is said {\sl ordered consistent}
if $h_1\leq h_2$ implies
$Tr(C_{h_1}\rho C_{h_1}^\ast)
\leq
Tr(C_{h_2}\rho C_{h_2}^\ast)$, where both $h_1$ and $h_2$
belong to two medium decohering families.
When all histories of a medium decohering family $\cal C$ are
ordered consistent, then $\cal C$ is said to be {\sl ordered consistent}.
Not even ordered consistency is individual.
Indeed, if we take ${\cal H}={\bf C}^2$ in example 1, then
$\cal C$ must be ordered consistent
with respect to $\rho$, but it does not with respect to $\rho_1$
and $\rho_2$ because it is not weakly decohering.
\vskip4.6mm
The lack of individuality exhibited by all these notions of
consistency is in striking contrast with the idea of consistency of
which they should be the mathematical representation. However, this
is not a problem for the logical coherence of the theories, but, rather,
it reflects their unability in implementing the individuality of
consistency.
\par
Furthermore, the fact that all notions of consistency so far proposed
are not individual gives rise to the suspect that individual
consistency is a {\sl chimera}.
\par
Now we show that on the contrary, at least for 2-events histories,
a meaningful notion of individual consistency exists,
which we call {\sl self-decoherence}. It is stronger than
medium decoherence. Furthermore, contrary inferences are forbidden
by self-decoherence.
\par
Our proposal is based on the concept of {\sl mirror projection} [7].
Given a 2-event history $h=(E_1,E_2)$ and a density operator
$\rho$, a projection operator $T$ is a mirror projection for
$(h,\rho)$ if
\par\noindent
M1. \quad
$[T,E_1]=[T,E_2]={\bf 0}$,
\par\noindent
M2. \quad
$Tr(TE_1\rho)=Tr(T\rho)=Tr(E_1\rho)$.
\par\noindent
To understand the physical meaning of the mirror projection, we notice
that, since (by (M1)) $T$ commutes with $E_1$, we may compute
the quantum conditional probabilities
$p(T\mid E_1)={Tr(TE_1\rho)\over Tr(E_1\rho)}$ and
$p(E_1\mid T)={Tr(TE_1\rho)\over Tr(T\rho)}$, which are both 1 because
of (M2). Therefore, the events
$T$ and $E_1$ are directly correlated: $T$ occurs iff $E_1$ occurs.
Given the history $h=(E_1,E_2)$ with $[E_1,E_2]\neq {\bf 0}$,
standard quantum theory is unable to describe the occurrence
of $h$.
The existence of a mirror projection $T$ for $(h,\rho)$ allows to
introduce the following notion of occurrence of $h$.
\par
{({\sl oc})}
{\sl the history $h$ occurs if both events $T$, which is
directly\par correlated to
$E_1$, and $E_2$ occur.}
\par\noindent
Then we are led to the following notion of consistency:
\par\noindent
D{\ninerm EFINITION} 3.
{\sl
A family $\cal C$ of 2-event histories is said self-decohering with respect to $\rho$
if there is a mirror projection for $(h,\rho)$, for all $h\in\cal C$.}
\vskip4.6mm\noindent
Interesting physical situations may be described by self-decohering
histories. The following example was suggested by some works
of M.O. Scully, B-G. Englert and H. Walter [8].
\par\noindent
E{\ninerm XAMPLE} 2. --
Let us consider the two-slits experiment for a particle which possesses,
besides the spatial degrees of freedom $(x_1,x_2,x_3)={\bf x}$,
an internal degree of freedom $s$ corresponding to a dichotomic
observable $S$ with spectrum $\sigma(S)=\{1,0\}$.
Such a system is described in the Hilbert
space $L_2({\bf R}^3)\otimes {\bf C}^2$. The event ``the particle goes
through slit 1 (resp., 2)'' is represented by the projection operator
$E_1$ (resp., $F_1$). Given any interval $\Delta$ on the final screen,
by $E_2$ we denote the projection operator which represents the event
``the particle hits the final screen in a point within $\Delta$''.
Therefore $h_1=(E_1,E_2)$ and $h_2=(F_1,E_2)$ are non-commutative histories
which generate a family $\cal C$. Now suppose that the state vector
of the particle is
$\Psi={1\over\sqrt 2}[\psi_1\otimes\mid 1\rangle+\psi_2\otimes\mid 0\rangle
]$, where $\psi_1$ (resp. $\psi_2$) is a spatial wave function
localized in slit 1 (resp., 2) when the particle is in the two-slits' region.
Therefore
$$
E_1\psi_1=\psi_1, \quad F_1\psi_2=\psi_2, \quad
E_1\psi_2= F_1\psi_1=0.
$$
In this situation the projection operators
$T=\vert 1\rangle\langle 1\vert$ and
$U=\vert 0\rangle\langle 0\vert$
are mirror projections for
$(h_1,\vert\Psi\rangle\langle\Psi\vert)$ and
$(h_2,\vert\Psi\rangle\langle\Psi\vert)$. Therefore the family $\cal C$
is self-decohering, so that the history $h_1$ (resp., $h_2$) may be interpreted as
``the particle hits the final screen in $\Delta$ passing through slit
1 (resp., 2)''.
Actually, the which-slit test can be performed for each individual sample
of the physical system
by measuring together
$E_2$, $T$ and $U$.
Then we assign history $h_1$ ($h_2$) to that sample
if both $E_2$ and $T$ ($U$) yield a positive outcome.
\vskip4.6mm
Now we prove that self-decoherence is an {\sl individual} property.
Let us suppose that (M2) holds for $\rho=\lambda\rho_1+(1-\lambda)\rho_2$.
From $Tr(E_1T\rho)=Tr(T\rho)$ we get
$$
\lambda Tr[(T-E_1T)\rho_1)]+(1-\lambda)Tr[(T-E_1T)\rho_2]=0.\eqno(3)
$$
The traces in this equation are non-negative because $E_1T\leq T$.
Therefore (3) implies
$Tr[(T-E_1T)\rho_1)]=Tr[(T-E_1T)\rho_2]=0$. In a similar way,
$Tr[(E_1-E_1T)\rho_1)]=Tr[(E_1-E_1T)\rho_2]=0$ follows from
$Tr(E_1T\rho)=Tr(E_1\rho)$.
Then $T$ must be a mirror projection for both $(h,\rho_1)$ and
$(h,\rho_2)$.
Thus individuality condition is satisfied by
self-decoherence.
\par
Now we prove that medium decoherence, and hence weak decoherence,
hold in a self-decohering family.
We limit ourselves to pure density operators $\rho=
\vert \psi\rangle\langle\psi\vert$: the extension to general density
operators is straightforward.
\vskip4.6mm\noindent
P{\ninerm ROPOSITION} 1.
{\sl
If $T$ and $U$ are mirror projections respectively for
$(h_1=(E_1,E_2),\rho)$, $(h_2=(F_1,E_2),\rho)$, where
$\rho=\vert\psi\rangle\langle\psi\vert$, then the following statement
holds.
$$
E_1\perp F_1\quad\hbox{implies}\quad
\langle\psi\mid E_1E_2F_1\psi\rangle=0.\eqno(4)
$$
}\noindent
P{\ninerm ROOF.}
Let $T$ and $U$ be mirror projections for
$(h_1,\rho)$ and $(h_2,\rho)$, respectively,
and let $T\lor U$ denote the projection operator which is the least
upper bound of $T$ and $U$.
If $E_1\perp F_1$, by
(M2) we get [9]
$$
T\psi\perp U\psi,\quad (T\lor U)\psi=T\psi+U\psi,
\quad
T\psi=(T\lor U)\psi-U\psi.
\eqno(5)
$$
Therefore,
$$
\eqalign{
\langle\psi\mid E_1E_2F_1\psi\rangle&=\langle T\psi\mid E_2U\psi\rangle
=\langle (T\lor U)\psi\mid E_2U\psi\rangle-\langle U\psi\mid E_2 U\psi
\rangle\cr
&=\langle\psi \mid (T\lor U)E_2 U\psi\rangle-\langle\psi\mid E_2U\psi\rangle\cr
&=\langle\psi \mid E_2(T\lor U)U\psi\rangle-\langle\psi\mid E_2U\psi\rangle\cr
&=\langle\psi \mid E_2U\psi\rangle-\langle\psi\mid E_2U\psi\rangle=0.\cr}
$$
In the fourth equation we have used the fact that since
$E_2$ commutes with both $T$ anf $U$, then $E_2$ must commute
with $T\lor U$ (see, for instance, theorem 2.24 in [10]).
Thus, proposition 1 is proved.
\vskip4.6mm
Individuality is not sufficient to assign the meaning of {\sl consistency}
to self-decoherence. A sensible notion of consistency should satisfy
conditions (C.0) and (C.1). Now, if $\cal C$ is self-decohering, the
probability of occurrence of $h=(E_1,E_2)\in\cal C$ which agrees with
({\sl oc}) is $p(E_1,E_2)=Tr(E_2T\rho )= Tr(E_2E_1\rho)$. Therefore, it satisfies
both (C.0) and (C.1). Furthermore, because of (M1) and (M.2) we have
$p(h)=Tr(E_2T\rho)=
Tr(E_2T\rho TE_2)=
Tr(E_2E_1\rho E_1E_2)=Tr(C_h\rho C_h^\ast)$.
Therefore we arrive at the same formula of the probability {\sl assumed}
by CHA, without imposing it. It turns out to be, rather, a natural consequence
of the notion of occurrence of a history ({\sl oc}) we have
introduced by means of the concept of mirror projection.
\vskip4.6mm
The possibility of contrary inferences is the main critique opposed
to CHA. Let us briefly
describe them. Suppose that ${\cal C}_1$ and ${\cal C}_2$ are
two {\sl different} weakly decohering families such that
$h_1=(E_1,E_2)\in{\cal C}_1$ and $h_2=(F_1,E_2)\in{\cal C}_2$, with
$E_1\perp F_1$. A. Kent [4] was able to find examples in which
the conditional probabilities $p_{{\cal C}_1}(h_1\mid E_2)={p_{{\cal C}_1}(h_1)\over p_{{\cal C}_1}(E_2)}$
and $p_{{\cal C}_2}(h_2\mid E_2)={p_{{\cal C}_2}(h_2)\over p_{{\cal C}_2}(E_2)}$ are both 1. Therefore,
when $E_2$ occurs we may state, according to CHA, that also $E_1$
occurs within the family ${\cal C}_1$, and that also $F_1$ occurs
within the family ${\cal C}_2$; on the other hand, $E_1\perp F_1$
means that the occurrence of $E_1$ excludes the occurrence of $F_1$:
then we have two inferences which are contrary to each other. They
do not entail logical inconsistency for CHA, because they take place
in {\sl different} consistent families.
But the meaning of the occurrence of $E_1$, or $F_1$,
once $E_2$ has occurred, becomes obscure. This
state of affairs has been judged negatively by some authors [4][11],
according to whom CHA is an unsatisfactory theory.
\par
We can easily prove that such kind of contrary inferences cannot take place
if we consider only self-decohering families.
Indeed, if ${\cal C}_1$ and ${\cal C}_2$ are self-decohering
we have
$$\eqalignno{
p_{{\cal C}_1}(h_1)+p_{{\cal C}_2}(h_2)
=&\langle\psi\mid E_1E_2E_1\psi\rangle
+\langle\psi\mid F_1E_2F_1\psi\rangle&\cr
=&\langle\psi\mid E_1E_2E_1\psi\rangle
+\langle\psi\mid F_1E_2F_1\psi\rangle
+&\cr
&+\langle\psi\mid E_1E_2F_1\psi\rangle
+\langle\psi\mid F_1E_2E_1\psi\rangle&{\rm by\hskip2mm prop.1}\cr
=&\langle\psi\mid (E_1+F_1)E_2(E_1+F_1)\psi\rangle
\leq\langle\psi \mid E_2\psi\rangle=p(E_2).&\cr}
$$
Then the sum of
$p_{{\cal C}_1}(h_1\mid E_2)={p_{{\cal C}_1}(h_1)\over p_{{\cal C}_1}(E_2)}$
and $p_{{\cal C}_2}(h_2\mid E_2)={p_{{\cal C}_2}(h_2)\over p_{{\cal C}_2}(E_2)}$
cannot be greater than 1. Thus, contrary inferences are forbidden.
\vskip4.6mm
We end with a necessary remark. At this stage we cannot state that
self-decoherence is the ultimate consistency's notion able to solve
all difficulties of CHA. Several questions should be seriously
examined. A problem is how to extend
the notion of self-decoherence to histories made up of more
than two events. Another question is whether the following
condition should be required for a property $\pi$ being
an individual property.
\item{C)}
{\sl If $\pi$ holds with respect to $\rho_1$ and $\rho_2$, then $\pi$
holds with respect $\lambda\rho_1+(1-\lambda)\rho_2$.}
\par\noindent
Actually, self-decoherence does not satisfy such further condition [12].
This notwithstanding, we think that self-decoherence
possesses sufficiently interesting features to be submitted to the
attention of researchers working in the field of foundations of
physics.
\vskip2pc\noindent
\item{[1]} 
R.B. Griffiths, J.Stat.Phys., {\bf 36}, 219 (1984).
\item{[2]}
R.B. Griffiths, Phys.Rev. {\bf A54}, 2759 (1996);
Phys.Rev. {\bf A57}, 1604 (1998),
\item{} R. Omn\`es, {\sl The interpretation of quantum mechanics},
Princeton Un. Press, Princeton 1994;
{\sl Understanding quantum mechanics},
Princeton Un. Press, Princeton 1999.
\item{[3]}
J.S. Bell, Rev.Mod.Phys., {\bf 38}, 447 (1966);
\item{}
S. Kochen and E.P. Specker, J.Math.Mech., {\bf 17}, 59 (1967).
\item{[4]}
A. Kent, Phys. Rev. Lett., {\bf 78} 2874 (1997);
Lect. Notes Physics, {\bf 559}, 93 (2000).
\item{[5]}
M. Gell-Mann and J.B. Hartle, Phys.Rev. {\bf D47}, 3345 (1993)
\item{[6]}
S. Goldstein and D.N. Page, Phys.Rev.Lett., {\bf 74}, 3715 (1995).
\item{[7]}
G. Nistic\`o and M.C. Romania, J.Math.Phys., {\bf 35}, 4534 (1994);
\item{} G. Nistic\`o, Found.Phys., {\bf 25}, 1757 (1995)
\item{[8]}
M.O. Scully, B-G. Englert and H. Walter,
Nature {\bf 351}, 111 (1991).
\item{[9]}
Here we prove the second statement of (5).
If we choose a basis $\{u_n\}$ of the Hilbert space $\cal H$
such that ${u_{1}=
{T\psi\over \Vert T\psi\Vert}}$
and ${u_{2}={U\psi\over \Vert U\psi\Vert}}$, then by representing
the vector $(T\lor U)\psi$ with respect to $\{u_n\}$ we have
$(T\lor U)\psi=T\psi+U\psi+P\psi$, where $P\psi\perp\{T\psi,U\psi\}$.
It is obvious that $(T\lor U)\leq (T+U)$.
Therefore $\langle\psi \mid (T+U)\psi\rangle=\Vert T\psi\Vert^2
+\Vert U\psi\Vert^2\ge
{\langle(T\lor U)\psi\mid (T\lor U)\psi\rangle}={\Vert T\psi\Vert}^2
+{\Vert U\psi\Vert}^2+{\Vert P\psi\Vert}^2$. Thus
$P\psi=0$ and $(T\lor U)\psi=T\psi+U\psi$.
\item{[10]}
C. Piron, {\sl Foundations of quantum physics},
Benjamin, Reading, Massachusetts 1976.
\item{[11]}
A. Bassi and G.C. Ghirardi, Phys. Lett. {\bf A257}, 247 (1999).
\item{[12]}
Indeed, if (C) holds, whenever $T_1$ and $T_2$ are mirror projections
for $(h,\vert\psi_1\rangle\langle\psi_1\vert$ and $(h,\vert\psi_2\rangle\langle\psi_2\vert$, then one mirror projection
$T$ should exist for both. This implies that such $T$ should be
a mirror projection for $(h,{\vert\psi_1+\psi_2\rangle\langle\psi_1+\psi_2\vert\over
\langle\psi_1+\psi_2\mid\psi_1+\psi_2\rangle}$. But this is not always true, as proved in [7].
\bye